\documentclass[journal,10pt,letterpaper]{IEEEtran}

\usepackage{graphicx}
\graphicspath{{Figures/}}
\DeclareGraphicsExtensions{.eps,.ps}
\usepackage{amsmath}
\usepackage{amssymb}

\begin{document}

\title{Proposal for a nanoscale variable resistor/electromechanical transistor}

\author{J.~B\"urki, C.~A.~Stafford, and D.~L.~Stein%
\thanks{\textit{This work has been submitted to the IEEE for possible
publication. Copyright may be transferred without notice,
after which this version will be superseded.}}
\thanks{Manuscript submitted July 9, 2008}%
\thanks{J.~B\"urki was with the Department of Physics, University of Arizona, 
  Tucson, AZ 85721.  
  He is currently with the Department of Physics and Astronomy, 
  California State University, Sacramento, 
  Sacramento, CA 95819-6041, USA (email: buerki@csus.edu)}
\thanks{C.~A.~Stafford is with the Department of Physics, 
  University of Arizona, 1118 E.\ Fourth Street, Tucson, AZ 85721 
  (email: stafford@physics.arizona.edu)}%
\thanks{D.~L.~Stein is with the Department of Physics and 
  Courant Institute of Mathematical Sciences, New York University, 
  New York, NY 10003 (email: daniel.stein@nyu.edu)}}


\maketitle

\begin{abstract}
  A nanoscale variable resistor consisting of a metal nanowire (active element), 
  a dielectric, and a gate, is proposed.  By means of the gate voltage, stochastic 
  transitions between different conducting states of the nanowire can be induced, 
  with a switching time as fast as picoseconds.  
  With an appropriate choice of dielectric, the transconductance of the device, 
  which may also be considered an ``electromechanical transistor,'' is shown 
  to significantly exceed the conductance quantum $G_0=2e^2/h$, a remarkable 
  figure of merit for a nanoscale device.  
\end{abstract}



\section{Background}
\label{sec:background}

Metal nanowires have attracted considerable interest in the past decade due
to their remarkable transport and structural properties \cite{Agrait03}.
Long gold and silver nanowires were observed to form spontaneously under
electron irradiation \cite{Kondo97,Rodrigues02b,Oshima03}, and appear to be
surprisingly stable.  Even the thinnest gold wires, essentially chains of
atoms, have lifetimes of the order of seconds at room temperature
\cite{Smit03a}.  
Metal nanowires exhibit striking correlations between their stability and
electrical conductance \cite{Urban04b,Mares07}.  
That these filamentary structures are stable at all is rather counterintuitive
\cite{Kassubek01,Zhang03}, but can be explained by electron-shell effects
\cite{Yanson99,Yanson01,Kassubek01,Zhang03,Burki03}.  

Because most of their atoms are at the surface, with low coordination numbers,
metal nanowires behave essentially like fluids~\cite{Zhang03}.
Classically, the Rayleigh instability would break up any wire whose length
exceeds its circumference~\cite{Kassubek01}.  
Nevertheless, nanowires violating the Rayleigh criterion have
been observed~\cite{Kondo97,Rodrigues02b,Oshima03}.  
The instability is suppressed through quantum effects, with stabilization 
occurring through the nanowire's electronic shell structure.  
A quantum linear stability analysis~\cite{Kassubek01,Urban03,Urban04,Urban06} 
showed the existence of ``islands of stability'' for discrete intervals of the radius~$R$. 
These correspond to conductance ``magic numbers'' that agree with those observed in
experiments.  For low enough temperatures, there remain finite regions
of~$R$ stable against long-wavelength perturbations.  Therefore, stable
wires exist only in the vicinity of certain ``magic radii'' and
consequently at quantized conductance values $G$ that are integer multiples
of the conductance quantum $G_0=2e^2/h$.

However, the linear stability analysis ignores large thermal fluctuations that
can lead to breakup of the wire.  Nanowire lifetimes are inferred from conductance
histograms, compiled by cycling a mechanically controllable break junction (MCBJ)
thousands of times~\cite{Yanson99,Yanson00,Yanson01}.  These studies indicate that
conductance peaks disappear above fairly well-defined temperatures.

B\"urki~{\it et al.\/}~\cite{BSS05} studied the lifetimes of these
nanowires using techniques developed in~\cite{MS01}.  By modelling thermal
fluctuations through stochastic Ginzburg-Landau classical field theories,
they constructed a self-consistent approach to the fluctuation-induced
``necking'' of nanowires that is in good agreement with
experiment~\cite{BSS05,BSS06}.  Their theory indicates that passivated
noble metal nanowires are sufficiently stable at room temperature to serve as
interconnects between nanoscale circuit elements.

\begin{figure}[b]
  \centering
  \includegraphics[width=0.95\columnwidth]{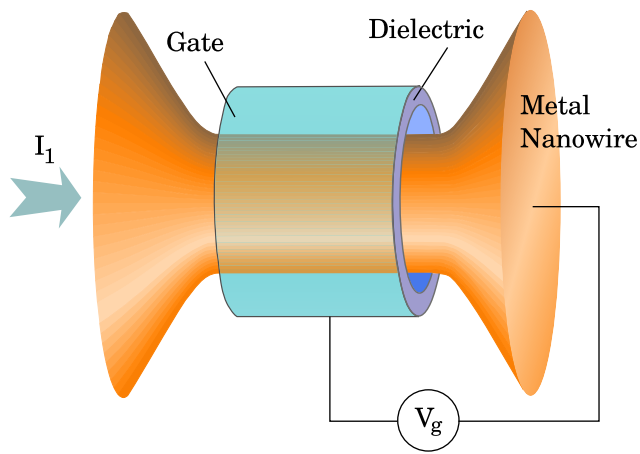}
  \caption{Diagram of the proposed device}
  \label{fig:diagram}
\end{figure}

Of particular interest for the applications considered in this proposal
is the nature of the barriers separating wires of different magic radii
(and hence quantized conductances).  These barriers can be surmounted in
several ways: among them are raising the temperature, applying strain,
shortening the wire, or changing the Fermi energy.  The first three are
discussed in~\cite{BSS05,BSS06}, but the last is new to this proposal.  For
the purposes of a new nanoscale device, the first two may be
unsatisfactory for various reasons having to do either with nonoptimal
operating conditions (temperature), or probable inability to implement
these controls on the nanoscale (strain).

On the experimental side, nanowires suitable for the proposed device, 
i.e.\ with conductance between a few and a few hundred conductance quanta, 
and lengths below or around a few nanometers have been fabricated 
using various techniques, including scanning tunnelling microscopy (STM)
\cite{Agrait93,Rubio96}, MCBJ \cite{Yanson98,Yanson99},
thin-film transmission electron microscopy (TEM) \cite{Kondo97,Kondo00},
electromigration \cite{Strachan05}, and electrochemical fabrication
\cite{Tao02c}.
Nanowires with diameters less than a nanometer have been directly 
observed \cite{Kondo97} using TEM to remain stable under low beam 
intensities below $5\,\text{A}/\text{cm}^2$ for the duration of observation.
Stochastic switching between different conductance values has been observed 
in contacts made using MCBJ \cite{Krans96}, while controllable switching 
has been achieved recently using electromigration to grow or shrink 
a nanobridge between two wires \cite{Terabe05}.  
A structural thinning process of the nanowire similar to the one described 
by the theory of B\"urki {\it et al.} \cite{BSS05,Burki07} has been observed 
to take place for gold nanowires in TEM experiments.  
The nanowire was observed \cite{Oshima03} to thin step by step via a process
where a structural step (corresponding to a change in radius of the order 
one atomic diameter) forms at one end of the wire and subsequently 
propagates along the wire.  

\section{Description of the device}
\label{sec:description}

Variable resistors are commonly used circuit elements in many electronic
applications.  However, their large size and slow response time have
heretofore limited their use primarily to the human-circuit interface.
In this article, we describe how the exploitation of quantum and
stochastic effects at the nanoscale allows one to
combine what would ordinarily be distinct macroscale circuit elements into
a single nanoscale device with multiple functionalities, and to achieve
response times on the order of picoseconds.

\begin{figure}[bt]
  \centering
  \includegraphics[width=0.95\columnwidth]{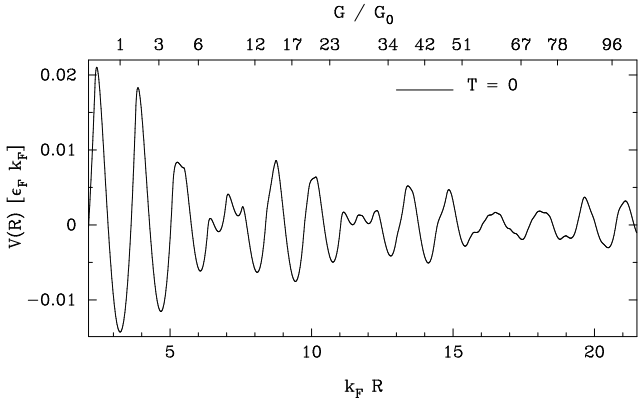}
  \caption{Electron-shell potential as a function of wire radius $k_F R$. The 
    conductance of the wires with ``magic radii'' is indicated on the top axis.
    From Ref.\ \cite{BSS05}}.
  \label{fig:shell_potential}
\end{figure}

The proposed device is illustrated in Fig.~\ref{fig:diagram}.  The physics 
behind its operation is the following: A metal nanowire is the active circuit 
element, and is embedded in a dielectric sheath, surrounded by an outer 
conductor of comparable dimensions, referred to as the gate.  
A positive/negative voltage applied to the gate enhances/depletes the density 
of carriers in the nanowire.  
Importantly, the resulting shift in the Fermi energy $E_F$ alters the 
electron-shell structure of the nanowire, which in turn determines its 
stability.

\begin{figure}[b]
  \centering
  \includegraphics[width=0.95\columnwidth]{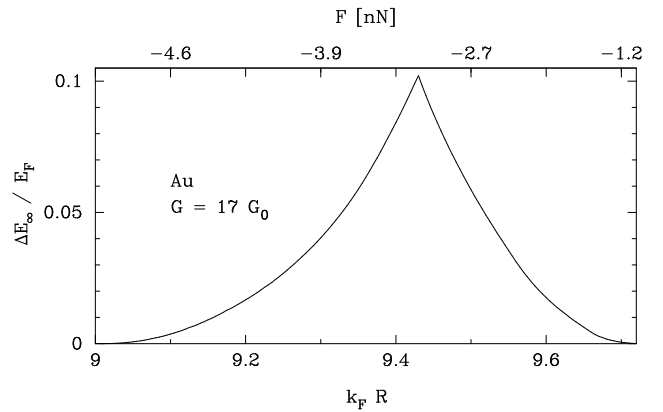}
  \caption{Escape barrier as a function of wire radius $k_FR$, or equivalently 
    applied stress $F$. Results correspond to a gold wire with a conductance 
    $G=17\,G_0$, following the calculations of Ref.\ \cite{BSS05}. }
  \label{fig:strain}
\end{figure}

The magic radii are the minima of the electron-shell potential (see
Fig.~\ref{fig:shell_potential}, and e.g.\ \cite{Burki03}), which depends 
on the dimensionless parameter $k_F R$, with $k_F$ the Fermi wavevector 
and $R$ the wire radius.  
A shift in $k_F$ is thus analogous to applying strain, and can be used to
induce rapid (i.e., on the scale of the Debye frequency) transitions
between neighboring magic radii.  These have conductances differing by $n
G_0$, where $n\geq 2$ is an integer (see Fig.~\ref{fig:shell_potential}).  (As a
rule of thumb, the jumps scale as $n\sim (\pi/4)k_F R$ for a wire with
initial radius $R$.)  The switching time between two adjacent magic radii
was shown~\cite{BSS05} to be given by the Kramers formula
\begin{equation}
\label{eq:Kramers}
\tau \sim \Gamma_0^{-1} \exp(\Delta E/k_B T),
\end{equation}
where $\Delta E$ is the energy barrier, $k_B$ is Boltzmann's constant, and
$T$ is the temperature.  The rate prefactor $\Gamma_0$, of order the Debye
frequency, was calculated explicitly in Ref.~\cite{BSS05}.  
The dependence of $\Delta E$ on the parameter $k_F R$ is illustrated in 
Fig.~\ref{fig:strain}.

The possibility of shifting $E_F$ electrostatically, as described above,
depends in an essential way on the crucial feature that the nanowire has a
radius of order nanometers, and thus has a very low density of states at
$E_F$.  As a function of the applied gate voltage~$V_g$, the shift
in $E_F$ is given by~\cite{Kassubek99}
\begin{equation}
\delta E_F = \frac{e V_g}{1 + (e^2/C_g)dN/dE}\, ,
\label{eq:dE_F}
\end{equation}
where $C_g$ is the mutual capacitance between gate and nanowire, and
$dN/dE$ is the density of states of the nanowire at $E_F$.  As discussed
in~\cite{Kassubek99}, the denominator in Eq.~(\ref{eq:dE_F}) can be well approximated
in terms of material and geometrical parameters, yielding a convenient
rule-of-thumb estimate
\begin{equation}
\delta E_F \approx \frac{e V_g}{1 + \alpha \, r_s \, \epsilon^{-1}G/G_0},
\label{eq:dE_F_est}
\end{equation}
where $r_s$ is the Fermi gas parameter for the nanowire material
(essentially the mean interelectron separation in the bulk metal), 
$\epsilon$ is the mean dielectric constant of the dielectric sheath,
and $\alpha$ is a dimensionless parameter of order unity, which depends
logarithmically on the device dimensions (Fig.~\ref{fig:diagram}).

In order to achieve the maximum switching speed, it is necessary to achieve
a shift $\delta (k_F R)\sim 1$ in the shell-potential parameter.  From
Eq.~(\ref{eq:dE_F_est}), this implies a preferred operating gate voltage
\begin{equation}
\frac{e V_g}{E_F} \sim \frac{\alpha \, r_s \, k_F R}{6 \, \epsilon}.
\label{eq:dVg_est}
\end{equation}
For typical metals, $r_s \sim 2$--3, while $k_F R \sim 10$ in the domain of
validity of the nanoscale free electron model~\cite{Stafford97,Burki05}.  
It is therefore desirable to use a dielectric with~$\epsilon \geq 10$ to 
minimize the necessary gate voltages.

\subsection{Transconductance}
\label{sec:transconductance}

Because the mechanical switching time of the nanoscale variable resistor
can be as short as picoseconds, it may also be thought of as an~{\em
electromechanical transistor\/}.  It is thus useful to compute its~{\em
transconductance}, a figure of merit used to characterize transistors.  The
transconductance $g_T$ can be estimated as
\begin{equation}
\label{eq:transconductance}
g_T=\frac{dI_1}{dV_g} \sim \frac{n \, G_0 V_{12} }{V_g}.
\end{equation}
Using Eq.\ (\ref{eq:dVg_est}) and $n\sim (\pi/4)k_F R$, one finds
\begin{equation}
\label{eq:trans2}
\frac{g_T}{G_0} \sim \frac{3\pi \epsilon}{2\alpha r_s} \frac{e V_{12}}{E_F}.
\end{equation}
For large dielectric constants~$\epsilon\geq 10$, and bias voltages~$V_{12}
\sim 1$V, one can thus achieve $g_T\gg G_0$, an exceptional figure of merit
for a nanoscale device~\cite{CSM06}, thereby enabling its advantageous use 
as an effective transistor.  
In addition to the structural switching time of order picoseconds, the 
electrical $RC$ rise time $\tau=C_g/G$ can be estimated to be of order 
1 femtosecond for typical device dimensions/materials, and so
is not a limiting factor in device performance.

\subsection{Ohmic$\leftrightarrow$non-Ohmic transition}
\label{subsec:ohmic}

\begin{figure}
  \centering
  \includegraphics[width=0.95\columnwidth]{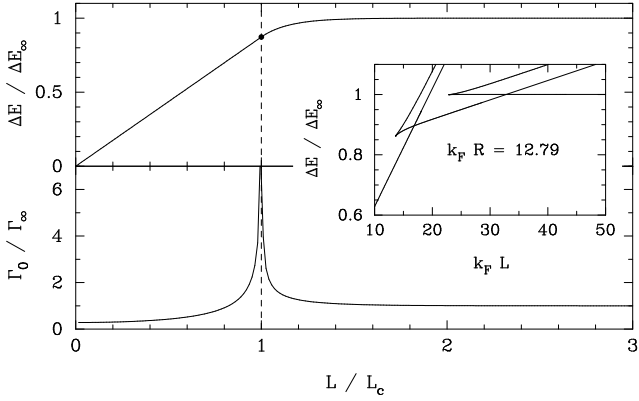}
  \caption{Escape barrier $\Delta E$ (top) and prefactor $\Gamma$ (bottom) 
    as a function of wire length, as calculated in \cite{BSS05}.  
    A second-order phase transition takes place 
    at the critical length $L_c$. For some wires, the transition is first-order 
    (inset).}
  \label{fig:transition}
\end{figure}

\begin{figure}[b]
  \centering
  \includegraphics[width=0.95\columnwidth]{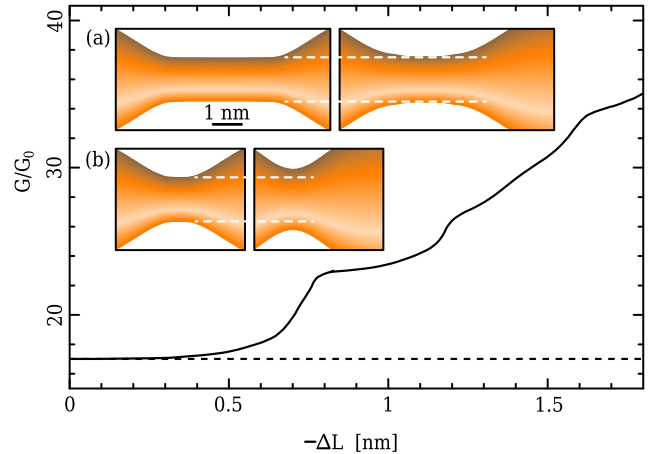}
  \caption{Conductance of a short (solid line) and long (dashed line) wire 
    under compression.  The initial and final shapes of the long (a) and 
    short (b) wires are shown in the inset.  From Ref.\ \cite{BSS06}.}
  \label{fig:IV}
\end{figure}

The device discussed above is one where barriers are controlled by shifting
the Fermi energy of the nanowire through electrostatic means.  Another
possibility is to change the wire length.  In~\cite{BSS05} it was predicted
that a transition in activation behavior occurs as a function of wire
length: below a critical length $L_c$, the barrier decreases rapidly with
length, while above it is roughly constant.  The transition can be
continuous (second-order) or discontinuous (first-order) (see
Fig.~\ref{fig:transition}).  This effect may have already been observed: a recent
study~\cite{YOT05} reported a transition from linear to nonlinear $I-V$
behavior in gold nanowires, as distance between electrodes shortened due to
applied bias.  In a Comment~\cite{BSS06}, we were able to explain this
result as a consequence of the transition in radius stability as a function
of wire length (cf.~Fig.~\ref{fig:transition}). 
This leads to another device
possibility, namely changing wire length directly by changing the applied
voltage.  This would convert a wire with linear $I-V$ characteristics to
one with nonlinear ones (see Fig.~\ref{fig:IV}).  
At the present time, it remains unclear how easily controllable such a 
transition might be.

\section{Fabrication of the device}
\label{sec:fabrication}

Commercial fabrication of a nanoscale variable resistor/electromechanical transistor
will require combining the three components of the device---metal nanowire, 
dielectric, and gate---integrated with other circuitry on a chip.  
Accordingly, the formation of the nanoscale
circuitry may be based on the following steps:

Step 1:
The initial structure of the nanowire, including its
electrical connection to the rest of the integrated circuit, can be formed 
with standard semiconductor fabrication and patterning processes such as 
for example E-beam direct write, or alternatively in the long term masked 
ion beam lithography, to deposit a metal wire tens of nanometers in diameter 
on the substrate (e.g., a Si wafer or other insulating substrate).  
Suitable metals for nanowire formation include Au, Ag, Cu, Pt, and Al, 
among others.

Step 2:
To form an active device, a short section of the nanowire must be thinned down 
to a diameter of order one nanometer using e.g.\ a focused
scanning electron microscope (SEM), electromigration \cite{Kim05,Saka05,Riveros06}, 
chemical etching \cite{Tao02c}, or a combination of these techniques.

Step 3:
The active segment of the nanowire must be encased in dielectric, which
also serves to passivate the nanowire surface, increasing durability.

Step 4: The nanowire
encased in dielectric must be placed in proximity to one or more metal gates, 
used to control the nanowire resistance through induced structural transitions.

\subsection{Dielectric}

To achieve optimal device characteristics, the space between the active segment 
of the nanowire and the gate(s) should be filled with a dielectric with 
$\epsilon \geq 10$.  
If a solid dielectric (only) is used, a small gap around the active segment of 
the nanowire must be provided (see Figure \ref{fig:pit}(b)) to permit the 
nanowire surface to fluctuate freely.  In that case, the mean dielectric constant of the
region between the nanowire and the gate(s) (including the gap) should exceed ten.  Many
intrinsic semiconductors could serve as suitable solid dielectrics with
$\epsilon \geq 10$ (e.g., Si, Ge, InSb, InAs, InP, GaSb, or GaAs).  The material should be
chosen so that the semiconducting energy gap exceeds the maximum desired voltage difference
between the gate and nanowire.

A liquid dielectric or combination of solid
and liquid dielectrics could also be utilized.
This would allow for optimal filling of the dielectric region, while still permitting free
motion of the nanowire surface.  
Liquid dielectrics have been used in conjunction with some of the previous
techniques \cite{Tao02a,Tao02b,Tao02c}, in the context of single
molecule measurements, as well as for STM measurements of metal contact
transport. In the latter context, they have been shown to have little influence
on the stability and transport properties of the nanocontact \cite{Tao02b}.

\begin{figure}[tb]
  \centering
  \includegraphics[scale=1]{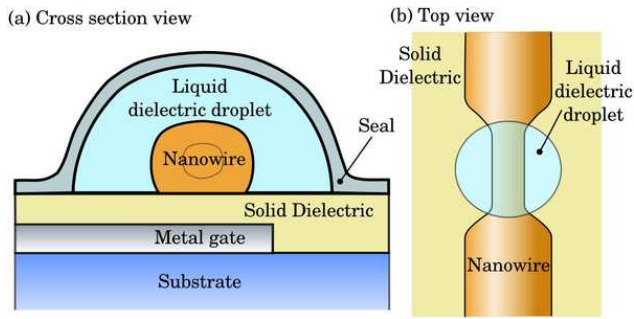}
  \caption{(a) Cross-section view and (b) top view diagrams of a device 
    with a nanodrop of liquid dielectric, discussed in Example 1. }
  \label{fig:droplet}
\end{figure}

Liquid dielectrics can have large dielectric constants and, being liquids,
would easily adapt to the shape of the nanowire, without preventing its
deformation.  They are thus ideal candidates for the proposed device.
For example, the dielectric constant of water at room temperature is close to 80.
While the finite conductivity of water might be problematic, deionized water still
has a dielectric constant of 15. As another example, Glycerol has a dielectric
constant above 40.
Various oil-based dielectrics [see e.g.\ US patent No 413189] have been developed
and could be suitable for the proposed device.
Alternatively, a dielectric gel or sol-gel or a combination of solid and gel dielectrics can 
be placed between the active segment of the nanowire and the gate(s).

\section{Examples}
\label{sec:examples}

Examples of specific nanowire devices are provided below.  It is important to note that
the device architectures described in the examples are generic in nature, 
and the list of examples included is not intended to be exhaustive.

\subsection*{Example 1}

In one embodiment of the proposed device (see Figure \ref{fig:droplet}),
a layer of solid dielectric is deposited on
a substrate prepatterned (using standard vapor deposition techniques)
with a metallic gate to address the nanowire device.  A metal
nanowire several tens of nanometers in diameter with a ``notch'' or constriction at the
desired location
is then deposited on the surface of the dielectric, in alignment
with the submerged gate electrode.
The nanowire at the notch can then be thinned down to the specified operating diameter
by electromigration, SEM, or chemical etching.
A nanodroplet of liquid dielectric is then deposited on the surface of the wafer,
immersing the nanowire device (see Figure \ref{fig:droplet}).
The nanowire device, together with the
droplet of liquid dielectric, is then hermetically sealed, e.g., with an epoxy seal.

\begin{figure}[bh]
  \centering
  \includegraphics[scale=1]{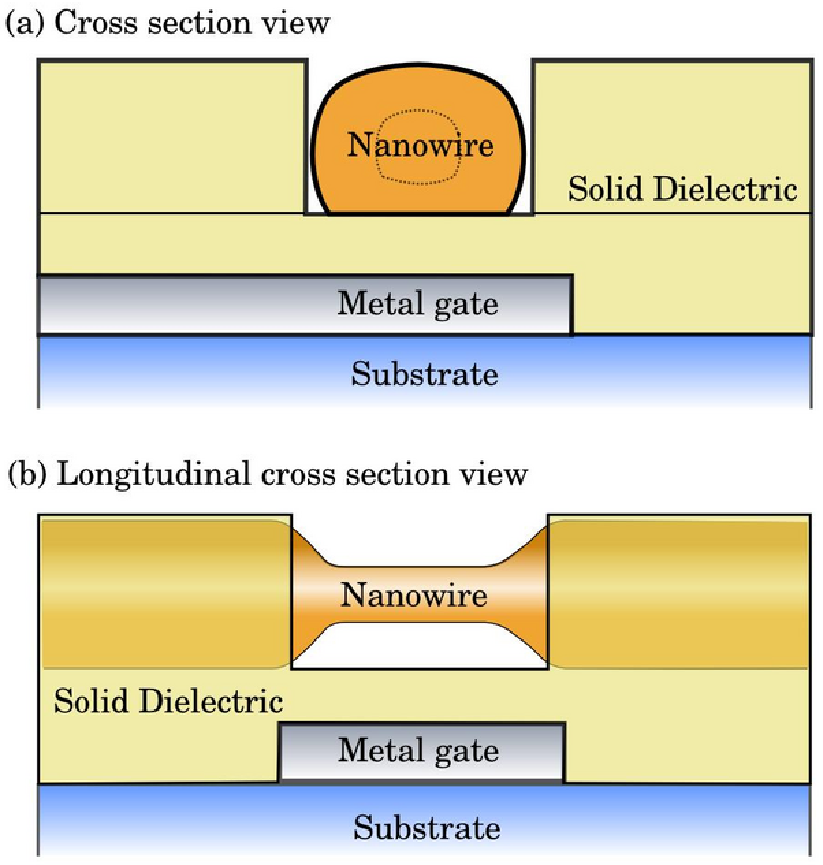}
  \caption{(a) Cross-section and (b) longitudinal cross-section diagrams of a device 
    with a solid dielectric, discussed in Example 2.  
    The gap shown around the nanowire may be filled with a liquid or gel dielectric 
    to improve performance.  
    Additionally, the cavity may be hermeticaly sealed, and a top gate can be added.}
  \label{fig:pit}
\end{figure}

\subsection*{Example 2}

In another embodiment of the proposed device (see Figure \ref{fig:pit}),
a layer of solid dielectric is deposited on a substrate prepatterned with a 
metallic gate to address the nanowire device, as in Example 1.  A metal
nanowire of uniform diameter several tens of nanometers
is then deposited on the surface of the dielectric, in alignment with the 
submerged gate electrode.  This fabrication step can be carried out within standard 
semiconductor patterning techniques, such as for example E-beam direct write or 
alternatively in the long term masked ion beam lithography. 
A further layer of solid dielectric is then deposited, fully encasing the nanowire.  
A nanoscale pit or cavity is then etched in the dielectric layer,
exposing the active segment of the nanowire.  The exposed segment of the nanowire 
is then thinned down to the specified diameter (of order one nanometer) via e.g., 
focused SEM, chemical etching, or electromigration, or a combination of these techniques. 

A hermetic seal can be applied to increase the durability of the nanowire device.
For example, an epoxy bubble seal may be used to enclose an inert
atmosphere (e.g., nitrogen or argon) about the exposed segment of the nanowire.  
Alternatively, a passivation layer over
the nanowire device is used to scavenge any small amounts of oxidant from 
the sealed environment. 

The pit containing the nanowire (see Figure \ref{fig:pit}(b)) can also be filled 
with a liquid or gel dielectric before the seal is applied, to enhance 
device performance.

For some applications, both a top gate (not shown) and a bottom gate are included, 
above and below the nanowire device, respectively.  
Multiple gates may be desirable e.g.\ to address individual devices in a large array.  
For example, if the gate voltage is chosen appropriately, the device will switch 
conducting states rapidly only if the voltage is applied to both gates.

\subsection*{Example 3}

\begin{figure}[bt]
  \centering
  \includegraphics[scale=1]{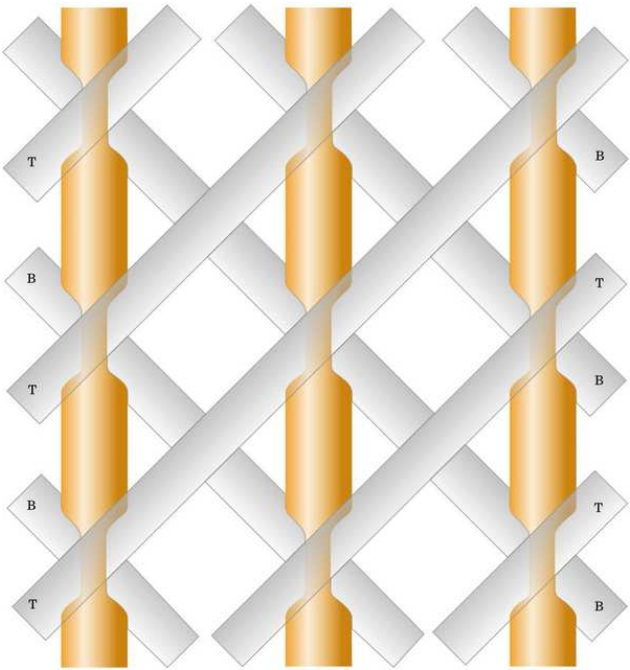}
  \caption{Top-view diagram of an array of devices with criss-crossing top (T) and 
    bottom (B) gates, discussed in Example 3. }
  \label{fig:array}
\end{figure}

In order to individually address individual devices in a large array on a chip, 
it may be desirable to fabricate a criss-crossing pattern of top and bottom gates 
(see Figure  \ref{fig:array}).
With an appropriate choice of operating voltages, only the nanowire device located 
at the intersection of the two active gates is addressed, and caused to switch 
conducting states.  

Because the three (or more) terminals of the nanoscale variable resistor are 
comprised of metal patterned by standard semiconductor fabrication techniques, 
such devices can be readily integrated with conventional circuitry on a chip.  
Because the throughput impedance of such a device is on the scale of several 
hundred to several thousand Ohms, appropriate amplification may be required 
to interface with standard CMOS circuitry.

\section*{Acknowledgments}
This work was supported by NSF Grant Nos.~0312028 (CAS) and PHY-0651077
(DLS).  
Part of this  work was done when CAS and DLS were at the Aspen Center for Physics.

\bibliographystyle{IEEEtran}
\bibliography{refs}

\begin{thebibliography}{10}
\providecommand{\url}[1]{#1}
\csname url@samestyle\endcsname
\providecommand{\newblock}{\relax}
\providecommand{\bibinfo}[2]{#2}
\providecommand{\BIBentrySTDinterwordspacing}{\spaceskip=0pt\relax}
\providecommand{\BIBentryALTinterwordstretchfactor}{4}
\providecommand{\BIBentryALTinterwordspacing}{\spaceskip=\fontdimen2\font plus
\BIBentryALTinterwordstretchfactor\fontdimen3\font minus
  \fontdimen4\font\relax}
\providecommand{\BIBforeignlanguage}[2]{{%
\expandafter\ifx\csname l@#1\endcsname\relax
\typeout{** WARNING: IEEEtran.bst: No hyphenation pattern has been}%
\typeout{** loaded for the language `#1'. Using the pattern for}%
\typeout{** the default language instead.}%
\else
\language=\csname l@#1\endcsname
\fi
#2}}
\providecommand{\BIBdecl}{\relax}
\BIBdecl

\bibitem{Agrait03}
N.~Agra{\"\i}t, A.~Levy~Yeyati, and J.~M. van Ruitenbeek, ``Quantum properties
  of atomic-sized conductors,'' \emph{Phys. Rep.}, vol. 377, pp. 81--279, 2003.

\bibitem{Kondo97}
Y.~Kondo and K.~Takayanagi, ``Gold nanobridge stabilized by surface
  structure,'' \emph{Phys. Rev. Lett.}, vol.~79, no.~18, pp. 3455--3458, 1997.

\bibitem{Rodrigues02b}
V.~Rodrigues, J.~Bettini, A.~R. Rocha, L.~G.~C. Rega, and D.~Ugarte, ``Quantum
  conductance in silver nanowires: Correlation between atomic structure and
  transport properties,'' \emph{Phys. Rev. B}, vol.~65, no.~15, p. 153402,
  2002.

\bibitem{Oshima03}
Y.~Oshima, Y.~Kondo, and K.~Takayanagi, ``High-resolution ultrahigh-vacuum
  electron microscopy of helical gold nanowires: junction and thinning
  process,'' \emph{J. Electron Microsc.}, vol.~52, pp. 49--55, 2003.

\bibitem{Smit03a}
R.~H.~M. Smit, C.~Untiedt, and J.~M. van Ruitenbeek, ``High-bias stability of
  monatomic chains,'' \emph{Nanotech.}, vol.~15, p. S472, 2004.

\bibitem{Urban04b}
D.~F. Urban, J.~B{\"u}rki, A.~I. Yanson, I.~K. Yanson, C.~A. Stafford, J.~M.
  van Ruitenbeek, and H.~Grabert, ``Electronic shell effects and the stability
  of alkali nanowires,'' \emph{Solid St. Comm.}, vol. 131, no. 9-10, pp.
  609--614, 2004.

\bibitem{Mares07}
A.~I. Mares, D.~F. Urban, J.~B{\"u}rki, H.~Grabert, C.~A. Stafford, and J.~M.
  van Ruitenbeek, ``Electronic and atomic shell structure in aluminum
  nanowires,'' \emph{Nanotechnology}, vol.~75, no.~26, p. 265403, 2007.

\bibitem{Kassubek01}
F.~Kassubek, C.~A. Stafford, H.~Grabert, and R.~E. Goldstein, ``Quantum
  suppression of the rayleigh instability in nanowires,'' \emph{Nonlinearity},
  vol.~14, pp. 167--177, 2001.

\bibitem{Zhang03}
C.-H. Zhang, F.~Kassubek, and C.~A. Stafford, ``Surface fluctuations and the
  stability of metal nanowires,'' \emph{Phys. Rev. B}, vol.~68, p. 165414,
  2003.

\bibitem{Yanson99}
A.~I. Yanson, I.~K. Yanson, and J.~M. van Ruitenbeek, ``Observation of shell
  structure in sodium nanowires,'' \emph{Nature}, vol. 400, pp. 144--146, 1999.

\bibitem{Yanson01}
A.~I. Yanson, J.~M. van Ruitenbeek, and I.~K. Yanson, ``Shell effects in alkali
  metal nanowires,'' \emph{Low Temp. Phys.}, vol.~27, pp. 807--820, 2001.

\bibitem{Burki03}
J.~B{\"u}rki, R.~E. Goldstein, and C.~A. Stafford, ``Quantum necking in
  stressed metallic nanowires,'' \emph{Phys. Rev. Lett.}, vol.~91, p. 254501,
  2003.

\bibitem{Urban03}
D.~F. Urban and H.~Grabert, ``Interplay of {R}ayleigh and {P}eierls
  instabilities in metallic nanowires,'' \emph{Phys. Rev. Lett.}, vol.~91, p.
  256803, 2003.

\bibitem{Urban04}
D.~F. Urban, J.~B{\"u}rki, C.-H. Zhang, C.~A. Stafford, and H.~Grabert,
  ``Jahn-teller distortions and the supershell effect in metal nanowires,''
  \emph{Phys. Rev. Lett.}, vol.~93, p. 186403, 2004.

\bibitem{Urban06}
D.~F. Urban, J.~B{\"u}rki, C.~A. Stafford, and H.~Grabert, ``Stability and
  symmetry breaking in metal nanowires,'' \emph{Physical Review B}, vol.~74, p.
  245414, 2006.

\bibitem{Yanson00}
A.~I. Yanson, I.~K. Yanson, and J.~M. van Ruitenbeek, ``Supershell structure in
  alkali metal nanowires,'' \emph{Phys. Rev. Lett.}, vol.~84, pp. 5832--5835,
  2000.

\bibitem{BSS05}
J.~B{\"u}rki, C.~A. Stafford, and D.~L. Stein, ``Theory of metastability in
  simple metal nanowires,'' \emph{Phys. Rev. Lett.}, vol.~95, p. 090601, 2005.

\bibitem{MS01}
R.~S. Maier and D.~L. Stein, ``Droplet nucleation and domain wall motion in a
  bounded interval,'' \emph{Phys. Rev. Lett.}, vol.~87, p. 270601, 2001.

\bibitem{BSS06}
J.~B{\"u}rki, C.~A. Stafford, and D.~L. Stein, ``Comment on {`N}onlinear
  current-voltage curves of gold quantum point contact{s'},'' \emph{Appl. Phys.
  Lett.}, vol.~88, p. 166101, 2006.

\bibitem{Agrait93}
N.~Agra{\"\i}t, J.~G. Rodrigo, and S.~Vieira, ``Conductance steps and
  quantization in atomic-size contacts,'' \emph{Phys. Rev. B}, vol.~47, pp.
  12\,345--12\,348, 1993.

\bibitem{Rubio96}
G.~Rubio, N.~Agra{\"\i}t, and S.~Vieira, ``Atomic-sized metallic contacts:
  mechanical properties and electronic transport,'' \emph{Phys. Rev. Lett.},
  vol.~76, pp. 2302--2305, 1996.

\bibitem{Yanson98}
A.~I. Yanson, G.~Rubio~Bollinger, H.~E. van~den Brom, N.~Agra{\"\i}t, and J.~M.
  van Ruitenbeek, ``Formation and manipulation of a metallic wire of single
  gold atoms,'' \emph{Nature}, vol. 395, pp. 783--785, 1998.

\bibitem{Kondo00}
Y.~Kondo and K.~Takayanagi, ``Synthesis and characterization of helical
  multi-shell gold nanowires,'' \emph{Science}, vol. 289, pp. 606--608, 2000.

\bibitem{Strachan05}
D.~R. Strachan, D.~E. Smith, D.~E. Johnston, T.-H. Park, M.~J. Therien, D.~A.
  Bonnell, and A.~T. Johnson, ``Controlled fabrication of nanogaps in ambient
  environment for molecular electronics,'' \emph{App. Phys. Lett.}, vol.~86, p.
  043109, 2005.

\bibitem{Tao02c}
H.~X. He, S.~Boussaad, B.~q. Xu, C.~Z. Li, and N.~J. Tao, ``Electromechanical
  fabrication of atomically thin metallic wires and electrodes separated with
  molecular-scale gaps,'' \emph{J. Electroanal. Chem.}, vol. 522, pp. 167--172,
  2002.

\bibitem{Krans96}
J.~M. Krans, J.~M. van Ruitenbeek, and L.~J. de~Jongh, ``Atomic structure and
  quantized conductance in metal point contacts,'' \emph{Physica B}, vol. 218,
  pp. 228--233, 1996.

\bibitem{Terabe05}
K.~Terabe, T.~Hasegawa, T.~Nakayama, and M.~Aono, ``Quantized conductance
  atomic switch,'' \emph{Nature}, vol. 433, pp. 47--50, 2005.

\bibitem{Burki07}
J.~B{\"u}rki, ``Discrete thinning dynamics in a continuum model of metallic
  nanowires,'' \emph{Physical Review B}, vol.~75, p. 205435, 2007.

\bibitem{Kassubek99}
F.~Kassubek, C.~A. Stafford, and H.~Grabert, ``Force, charge, and conductance
  of an ideal metallic nanowire,'' \emph{Phys. Rev. B}, vol.~59, no.~11, pp.
  7560--7574, 1999.

\bibitem{Stafford97}
C.~A. Stafford, D.~Baeriswyl, and J.~B{\"u}rki, ``Jellium model of metallic
  nanocohesion,'' \emph{Phys. Rev. Lett.}, vol.~79, pp. 2863--2866, 1997.

\bibitem{Burki05}
J.~B{\"u}rki and C.~A. Stafford, ``On the stability and structural dynamics of
  metal nanowires,'' \emph{Applied Physics A}, vol.~81, no.~8, pp. 1519--1525,
  2005.

\bibitem{CSM06}
D.~M. Cardamone, C.~A. Stafford, and S.~Mazumdar, ``Controlling quantum
  transport through a single molecule,'' \emph{Nano Letters}, vol.~6, p. 2422,
  2006.

\bibitem{YOT05}
M.~Yoshida, Y.~Oshima, and K.~Takayanagi, ``Nonlinear current-voltage curves of
  gold quantum point contacts,'' \emph{Appl. Phys. Lett.}, vol.~87, p. 103104,
  2005.

\bibitem{Kim05}
W.~J. Kim and M.~N. Carr, S. M.~Wybourneb, ``Direct contact buckling of
  electrochemically grown gold nanowires,'' \emph{Appl. Phys. Lett.}, vol.~87,
  p. 173112, 2005.

\bibitem{Saka05}
M.~Saka and R.~Ueda, ``Formation of metallic nanowires by utilizing
  electromigration,'' \emph{J. Mater. Res.}, vol.~20, no.~10, p. 2712, 2005.

\bibitem{Riveros06}
G.~Riveros, S.~Green, A.~Cortes, H.~G{\'o}mez, R.~E. Marotti, and E.~A.
  Dalchiele, ``Silver nanowire arrays electrochemically grown into nanoporous
  anodic alumina templates,'' \emph{Nanotechnology}, vol.~17, p. 561, 2006.

\bibitem{Tao02a}
B.~Xu, H.~Huixin, and N.~J. Tao, ``Controlling the conductance of atomically
  thin metal wires with electrochemical potential,'' \emph{J. Am. Chem. Soc.},
  vol. 124, pp. 13\,568--13\,575, 2002.

\bibitem{Tao02b}
H.~X. He, C.~Shu, C.~Z. Li, and N.~J. Tao, ``Adsorbate effect on the mechanical
  stability of atomically thin metallic wires,'' \emph{J. Electroanal. Chem.},
  vol. 522, pp. 26--32, 2002.

\end{thebibliography}


\begin{IEEEbiographynophoto}{J\'er\^ome B\"urki}
received the Ph.D. degree in physics from the University of Fribourg, 
Switzerland, in 2000.  
From 2004 to 2008, he held a position of Research Associate in the 
Physics Department at the University of Arizona.  
He is currently Assistant Professor of Physics in the Department 
of Physics and Astronomy at California State University, Sacramento.  
His research focuses on quantum effects on various properties, 
including nonlinear dynamics, of metal nanostructures.  
\end{IEEEbiographynophoto}

\begin{IEEEbiographynophoto}{Charles A. Stafford}
received the Ph.D.\ degree in physics from Princeton University, 
Princeton, New Jersey, in 1992.  
Since 1998, he has been on the faculty of the Physics Department 
at the University of Arizona, where he is currently Associate 
Professor of Physics.  
His research has focused on strongly-correlated electron systems and, 
more recently, on quantum transport and cohesion in nanostructures.  
His current interests include electron transport in single-molecule 
heterojunctions and the structural dynamics of metal nanowires.  
\end{IEEEbiographynophoto}

\begin{IEEEbiographynophoto}{Daniel L. Stein}
received the Ph.D.\ degree in physics from Princeton University, 
Princeton, New Jersey, in 1979.  Since 2005, he has been a Professor 
of Physics and Mathematics at New York University.  
His research in recent years has focused primarily on quenched disorder 
in statistical mechanical systems, particularly spin glasses, and 
on stochastic processes in both macroscopic and nanoscale systems.
\end{IEEEbiographynophoto}
\vfill

\end{document}